# Spin Waves as Metric in a Kinetic Space-Time

Lukas A. Saul

University of New Hampshire

March, 2003

**Abstract**

1) A wave equation is derived from the kinetic equations governing media with rotational as well as translational degrees of freedom. In this wave the fluctuating quantity is a vector, the bulk spin. The transmission is similar to compressive waves but propagation is possible even in the limit of incompressibility, where such disturbances could become dominant. 2) In this context a kinetic theory of space-time is introduced, in which hypothetical constituents of the space-time manifold possess such a rotational degree of freedom (spin). Physical fields (i.e. electromagnetic or gravitational) in such a theory are represented as moments of a statistical distribution of these constituents, as in the techniques of fluid mechanics. The spin wave equation from 1) is treated as a candidate for governing light and metric. Such a theory duplicates to first order Maxwell's equations of electromagnetism, Schrödinger's equation for the electron, and the Lorentz transformations of special relativity. Slight deviations from the classical approach are predicted and should be experimentally verifiable.

## I. SPIN – KINETICS

### A. The Distribution Function and Transport Equations

A statistical description of ensembles of rigid bodies is useful in many contexts, for example molecules as constituents of a fluid or gas, or grains in a plasma. In a detailed series of papers [Curtiss, 1957], a modified Boltzmann equation and transport equations for such a distribution were derived, using a distribution $\tilde{f} = \tilde{f}(\vec{x}, \vec{v}, \vec{\alpha}, \vec{\omega}; t)$ in which particles have only positional, translational, orientational, and rotational freedom in time. We follow a similar approach here, with a simplified distribution, assuming uniformity in the orientational distribution of the particles, i.e. assuming for now that $f = f(\bar{x}, \bar{v}, \bar{\omega}; t)$ fully characterizes the system. Here $\bar{x}$ is the position, $\bar{v}$ is the velocity, and $\bar{\omega}$ is the angular velocity of a constituent particle, at a time $t$. The distribution function gives the positive ten dimensional phase space density of particles with these properties, and the Boltzmann equation describing its evolution is (no external forces):

$$\frac{\partial f}{\partial t} + \frac{\partial f}{\partial x_i} v_i = \left\{\frac{\partial f}{\partial t}\right\}_{col.} \quad (1)$$

From this equation, we can integrate over all phase space ($d^6\Gamma$ or $d^3\bar{v}d^3\bar{\omega}$), to obtain the usual continuity equation:

$$\frac{\partial n}{\partial t} + \frac{\partial}{\partial x_i}(n\bar{v}_i) = 0 \quad (2)$$

where we have assumed that particles are neither created nor destroyed in collisions, and we have defined the number density $n(\bar{x}, t) = \int f d^6\Gamma$. We can also multiply (1) by $\bar{v}$ or by $\bar{\omega}$ before integrating, giving the transport equations:

$$\frac{\partial}{\partial t}(n\bar{v}_i) + \frac{\partial}{\partial x_j}(n\langle v_i v_j \rangle) = -E_i \quad (3)$$

$$\frac{\partial}{\partial t}(n\bar{\omega}_i) + \frac{\partial}{\partial x_j}(n\langle \omega_i v_j \rangle) = F_i \quad (4)$$

The notation is that: $\langle Q \rangle = \frac{1}{n}\int Q f d^6\Gamma$ is a function of space and time only, for any function Q. For ease of notation we write averages $\bar{v}_i = \langle v_i \rangle$ and similarly for $\omega$. We have also defined two vectors:

$$-E_i = \left\{\frac{\partial n\bar{v}_i}{\partial t}\right\}_{col} = \int \left\{\frac{\partial f}{\partial t}\right\}_{col} v_i d^6\Gamma \quad (5)$$

$$F_i = \left\{\frac{\partial n\bar{\omega}_i}{\partial t}\right\}_{col} = \int \left\{\frac{\partial f}{\partial t}\right\}_{col} \omega_i d^6\Gamma \quad (6)$$

representing the collision terms in the above conservation equations. In non rotational media, E will vanish as linear momentum is neither created nor destroyed in the (fully elastic) collisions.

It is useful to represent the velocity and rotation of a particle as a sum of the bulk speed and any 'peculiar' motion, i.e.: $v_i = \bar{v}_i + V_i$ and $\omega_i = \bar{\omega}_i + \Omega_i$ (capital letters represent the peculiar motion). The convection terms then can be defined:

$$n\langle v_i v_j \rangle = n\bar{v}_i \bar{v}_j + n\langle V_i V_j \rangle \equiv n\bar{v}_i \bar{v}_j + p_{ij} \qquad (7)$$

$$n\langle \omega_i v_j \rangle = n\bar{\omega}_i \bar{v}_j + n\langle \Omega_i V_j \rangle \equiv n\bar{\omega}_i \bar{v}_j + s_{ij} \qquad (8)$$

This spin pressure tensor ($s_{ij}$), will vanish if the distribution function is isotropic in $\vec{V}$ or $\vec{\Omega}$, just as the off-diagonal components of the pressure tensor ($p_{ij}$) do for f isotropic in $\vec{V}$, as the integrand will be odd in these variables. Also the collision terms for the velocity equation and the angular velocity equation $\vec{E}$ and $\vec{F}$ will vanish if $\left\{\dfrac{\partial f}{\partial t}\right\}_{col}$ is isotropic in $\vec{V}$ or $\vec{\Omega}$ respectively.

## B. Wave Equations

The first oscillations usually discussed in the kinetic theory of gases are compressive or sound waves. Equations (2) and (3) can be combined into one, by taking the time derivative of the first and the spatial gradient of the other, yielding:

$$\frac{\partial^2 n}{\partial t^2} = \frac{\partial}{\partial x_i}\frac{\partial}{\partial x_j}(n\langle v_i v_j \rangle) + \frac{\partial E_i}{\partial x_i} \qquad (9)$$

This equation governs compressive or sound-like waves. It reduces to the ordinary form for $E_i$ vanishing or constant, the pressure tensor isotropic and static, and the bulk speed vanishing or constant.

Another wave equation emerges from transport of the spin tensor. Multiplying the master equation (1) by the tensor $v_i \omega_j$ and integrating over phase space yields the transport equation:

$$\frac{\partial}{\partial t}(n\langle\omega_i v_j\rangle) + \frac{\partial}{\partial x_k}(n\langle\omega_i v_j v_k\rangle) = G_{ij} \qquad (10)$$

where the collision term $G_{ij}$ is a tensor defined analogously with the vectors $\bar{E}$ and $\bar{F}$. The rank three tensor moment in the second term is the *spin pressure transport tensor*.

We can now take the derivative of (10) with respect to $x_j$, and the derivative of (4) with respect to time, and combine these two equations:

$$\frac{\partial^2(n\bar{\omega}_i)}{\partial t^2} = \frac{\partial}{\partial x_j}\frac{\partial}{\partial x_k}(n\langle\omega_i v_j v_k\rangle) + \frac{\partial F_i}{\partial t} - \frac{\partial G_{ij}}{\partial x_j} \qquad (11)$$

At this point we have made no assumptions of the ten dimensional distribution function as defined, other than its statistical applicability. To see the power of this wave equation, consider the simplest case. Assume appropriate isotropy of the distribution function collision term so that the source terms, the vector F and tensor G, vanish. Take a frame of reference moving with the bulk flow so that $\bar{v}_i = 0$ for i=0,1,2. Equation (11) can now be written:

$$\frac{\partial^2(n\bar{\omega}_i)}{\partial t^2} = \frac{\partial}{\partial x_j}\frac{\partial}{\partial x_k}\left[n\bar{\omega}_i\langle V_j V_k\rangle + n\langle\Omega_i V_j V_k\rangle\right] \qquad (12)$$

If we further assume the distribution function to be isotropic in peculiar angular velocity $\bar{\Omega}$, the final moment will vanish. We can now assume constant density in space and time and cancel $n$ from this equation, giving the form:

$$\frac{\partial^2\bar{\omega}_i}{\partial t^2} = \frac{\partial}{\partial x_j}\frac{\partial}{\partial x_k}\left[\bar{\omega}_i\langle V_j V_k\rangle\right] \qquad (13)$$

At this point we have a familiar wave equation. An isotropic pressure tensor $\langle V_j V_k\rangle = c^2\delta_{jk}$ gives the standard form

$$\frac{\partial^2\bar{\omega}_i}{\partial t^2} = c^2\nabla^2\bar{\omega}_i, \qquad (14)$$

with d'Alembert solutions propagating at the speed c.

Such a wave equation would govern propagation of bulk spin gradients in incompressible kinetic media with rotational freedom. Presumably wave power would be added to these modes in nearly incompressible fluids. We now proceed to look at another potential arena where such fluctuations could exist, and examine their consequences.

## II. KINETIC SPACE-TIME
### A. Concept and Motivation

In kinetic theory, macroscopic physical properties (i.e. temperature and density) are attributed to statistical dynamic interactions of microscopic constituents (i.e. molecules). This method can be applied to physical fields present in a vacuum by postulating the existence of basic constituents of space-time itself. In this framework, we can attempt to explain such physical observables as the electromagnetic field vectors, quantum wave-functions, and the space-time metric as moments of a statistical distribution. Some authors have suggested names for such fundamental entities arising in related theories, using terms for fundamental atoms of space such as *monads*, *inertons*, or *Planck masses*. We consider here the model of interacting non-spherical particles making up the manifold of space-time [Meno 1991]. We follow Meno and call them *gyrons*.

The idea of physical observables as moments or averages of some function will not be unfamiliar to the quantum mechanic. The vacuum has fluctuations and energies, measured experimentally with the Casimir effect, something difficult to explain if a vacuum is truly "empty". The *zitterbewegung* quantum motion of electrons first discussed by Schrödinger in interpreting the Dirac equation [see e.g. Barut & Bracken 1981] can be viewed as Brownian motion of gyrons.

Indeed, the motivation for a kinetic theory of space-time largely comes from explaining such quantum phenomena as vacuum fluctuations and wave-functions. The Madelung transformation demonstrated symmetry of non-relativistic quantum mechanics and fluid mechanics. In particular the continuity and momentum equations for irrotational flow $(\nabla \cdot \bar{v} = 0)$ can be tranformed by introducing a velocity potential S and making the substitution [Madelung, 1926]:

$$\psi = \sqrt{n} e^{iS} \quad \text{where} \quad \bar{v}_i = \frac{\hbar}{m} \frac{\partial S}{\partial x_i}. \quad (14)$$

This transformation turns equations (2) and (3) into the complex Schrödinger equation,

$i\hbar \frac{\partial \psi}{\partial t} = \frac{-1}{2m} \nabla^2 \psi + V\psi$, for an appropriately defined quantum potential V [e.g. Winterburg, 1995]. It has also been suggested that in such a Madelung fluid, quantum behavior is a consequence of a fundamental existence of spin [Recami & Salesi, 1995], who also derive the quantum potential from hydrodynamic variables. Recent work has showed that the complex quantum mechanical operators emerge naturally as moments of a distribution function under such a transform [Kaniadakis, 2001].

Motivation for a kinetic theory of space-time is not only from the quantum realm. Classical space-time is endowed with inherent characteristics as well. The electromagnetic field and the gravitational potential (or space-time curvature) permeate all space, even in a vacuum. An atomistic or kinetic space-time would eliminate the troublesome "action at a distance" aspect of these forces and could explain the fields in terms of hypothetical internal motions.

Indeed, fluid explanations of electromagnetic phenomena have been used since the first observations of electromagnetic effects. We follow [Marmanis, 1998], re-deriving these symmetries based on the kinetic space-time theory.

### B. Symmetries and Electromagnetism

The observed constancy of distances in space suggests a dense, nearly incompressible distribution of the hypothetical gyrons. In such a medium, collisions must be commonplace. In this scenario, a bulk spin $\bar{\bar{\omega}}$ will be quickly coupled to the vorticity, defined as $\nabla \times \bar{\bar{v}}$. If we define proportional variables:

$$B_i \sim n\bar{\omega}_i, \quad \text{and} \quad A_i \sim n\bar{v}_i, \quad (15)$$

with appropriate constants, this coupling can be written in a form analogous to the definition of the magnetic vector potential: $\bar{B} = \nabla \times \bar{A}$. Using these new variables, and assuming an isotropic pressure tensor $\langle V_i V_j \rangle = c^2 \delta_{ij}$, equation (3) can be written as the definition of the electric field:

$$\vec{E} = -\nabla\phi - \frac{\partial \vec{A}}{\partial t},$$ where we have defined $\phi \sim nc^2$. (16)

The reason for the negative sign for the collision term in (3) is now apparent, and can be traced back to Ben Franklin. The symmetry between the first transport equations and electromagnetism becomes more compelling when we see that the continuity equation (2) can be written as the Lorentz condition: $\frac{\partial \phi}{\partial t}\frac{1}{c^2} + \nabla \cdot \vec{A} = 0$ [Meno, 1991].

Armed with these definitions, the wave equation (13) is seen to govern fluctuations of the magnetic field, and the coupling to vorticity, which gives $\nabla \cdot \vec{B} = 0$, limits the fluctuations of this field to be transverse to the direction of propagation ($\vec{k} \cdot \vec{B} = 0$). This is precisely what is observed in electromagnetic radiation. The coupling of spin to vorticity further tells us that $\vec{F} \sim -\nabla \times \vec{E}$, and so a vanishing spin pressure tensor in (4) gives $\frac{\partial \vec{B}}{\partial t} = -\nabla \times \vec{E}$. Maxwell's equations emerge naturally, with the transformations (15,16), and electric charge taken to be defined as a proportional to the divergence of the collision term $\vec{E}$.

We also note that this interpretation of the electromagnetic field gives a physical interpretation of the Lorentz force. A frame moving perpendicular to the magnetic field (bulk spin) will see collisions preferentially on the leading side of gyrons with which it collides, creating an electric field (bulk velocity collision term) $\vec{E} \sim \vec{\bar{v}} \times \vec{B}$.

### III. METRIC
#### A. Flat Space-Time

The Special Theory of Relativity (SR) firmly placed light as the definer of metric, with the postulate that the speed of light is the same in all reference frames. This brilliant choice enabled our current definition of a meter as the distance traveled by light in a vacuum in 1/299,792,458 of a second, and did not require a universal rest frame or absolute time. It also makes sense because the forces that hold together material bodies are electromagnetic. A metric based on sizes of material (chemical) bodies such as meter sticks or human beings must be inherently electromagnetic. Because the wave equation is not invariant under the Galilean transformation, this choice requires that the Lorentz transform be used for frames moving with different velocities, and the equations of motion modified according to SR [Einstein, 1905].

In the kinetic theory of space-time, the gyrons are assumed to move in perfect Euclidean coordinate space, in which the transformation of Galileo is the proper one to change reference frames. However, this space is inaccessible to observations, as all observable activity takes place on the framework of the space-time created by the gyrons, with light waves defining the metric. With some assumptions of the distribution function of the gyrons, we attempt to examine the light wave equation and determine an appropriate form for this space-time metric.

Returning to the general wave equation (11), we find in the absence of the collision/source terms F and G, assuming constant density $n$: $\frac{\partial^2 \bar{\omega}_i}{\partial t^2} = \frac{\partial}{\partial x_j} \frac{\partial}{\partial x_k} \langle \omega_i v_j v_k \rangle$. This cannot be solved without a set of functions, or kinetic equations of state, of the form:

$$\langle \omega_i v_j v_k \rangle \equiv f_{ijk}(\bar{\omega}_i) \quad \text{(no sum)}. \qquad (17)$$

For the case of small fluctuations $\delta \bar{\omega}_i$, these functions can then be taken as linear: $f_{ijk}(\delta \bar{\omega}_i) \cong \delta \bar{\omega}_i \cdot g_{jk}$, with the $g_{jk}$ constants that can come out of the derivatives in the wave equation. In Minkowski flat space-time (in vacuum) the distribution is isotropic in peculiar rotation and peculiar velocity, and the equation of state is:

$$f_{ijk}(\bar{\omega}_i) = \bar{\omega}_i \langle V_j V_k \rangle = \bar{\omega}_i \delta_{jk} c^2. \qquad (18)$$

This gives the wave equation the standard form (14), and the light sphere at a time dt from a pulse at the origin is described in coordinate form by $dx_1^2 + dx_2^2 + dx_3^2 = c^2 dt^2$. The Lorentz transformation leaves this sphere invariant and in more generality also keeps invariant the interval: $ds^2 = dx_1^2 + dx_2^2 + dx_3^2 - c^2 dt^2$. The Minkowski geometry emerges from our choice of spin waves as metric.

### B. Curved Space-Time

If we consider a small isotropic perturbation of the equation of state in space, we can rewrite the components of the metric function:

$$f_{ijk}(\bar{\omega}_i) = c^2(1-\eta)\delta_{jk}\bar{\omega}_i \quad \text{with} \quad \eta = \eta(\bar{x},t) \text{ small.} \quad (19)$$

The wave equation becomes: $\dfrac{\partial^2 \bar{\omega}_i}{\partial t^2} = c^2(1-\eta)\nabla^2 \bar{\omega}_i$, which gives an invariant interval: $ds^2 = dx_1^2 + dx_2^2 + dx_3^2 - c^2(1-\eta)dt^2$. In the usual form of the four dimensional metric tensor $ds^2 = g_{\mu\nu}x^\mu x^\nu$ we can recognize the identification $g_{00} = (1-\eta)$, which returns Newtonian gravity if $\eta = 2\Phi/c^2$ where $\Phi$ is the classical gravitational potential $MG_N/r$, and $G_N$ is Newton's gravitational constant [e.g. Dirac 1975].

However, a gravitational source region will not necessarily perturb the spin pressure transport tensor isotropically. To include effects of a possible first order anisotropy on the space-time metric, we start with the wave equation (13), for which the equations of state are $f_{ijk}(\bar{\omega}_i) = \bar{\omega}_i \langle V_j V_k \rangle$. We again assume the distribution function is isotropic in peculiar translation and spin of the gyrons. However, we do not assume it is isotropic in $V_i^2$, in other words we take the gyron pressure tensor diagonal such that the three components are not necessarily equal. Assume the pressure tensor has two independent components, parallel to the radial vector and perpendicular to it, i.e.:

$$f_{ijj}(\bar{\omega}_i) = \bar{\omega}_i c^2(1-\eta) \text{ for } j \neq r \quad \& \quad f_{irr}(\bar{\omega}_i) = \bar{\omega}_i c^2 \quad (i = r, \phi, \theta). \quad (20)$$

With these assumptions, (13) becomes (in spherical coordinates, and factoring out $(1-\eta)$ on the right hand side):

$$\frac{\partial^2 \bar{\omega}_i}{\partial t^2} = c^2(1-\eta)\left[\frac{1}{(1-\eta)}\frac{\partial^2 \bar{\omega}_i}{\partial r^2} + r^2\frac{\partial^2 \bar{\omega}_i}{\partial \phi^2} + r^2 \sin^2\theta \frac{\partial^2 \bar{\omega}_i}{\partial \theta^2}\right] \quad (21)$$

The light sphere or ovoid produced by a point source in a time dt in this geometry is described by $dr^2(1-\eta)^{-1} + r^2 d\phi^2 + r^2 \sin^2\theta d\theta^2 = c^2(1-\eta)dt^2$. The general invariant space-time interval is:

$$ds^2 = dr^2(1-\eta)^{-1} + r^2 d\phi^2 + r^2 \sin^2\theta d\theta^2 - c^2(1-\eta)dt^2, \quad (22)$$

which is the Schwarzschild solution to Einstein's equations for a perturbation $\eta = 2\Phi/c^2$. This line element has been shown to predict the precession of Mercury's orbit and is consistent with all observations of orbital motion and the gravitational redshift [e.g. Dirac 1975].

It is also possible to apply this theory to other dynamic metrics such as cosmological models. The Friedmann line element, for example, could be viewed as a universe in which the gyron pressure is steadily changing, creating the cosmological redshift and observed expansion of the universe.

### III. CONSEQUENCES AND PREDICTIONS

While the symmetries suggested by a kinetic theory of space-time and its more physical interpretation of forces are compelling, the *post*diction of Maxwell's equations and SR are not. What is needed is a concrete *pre*diction to test the theory, and place it in a scientifically verifiable framework. Two such ways to experimentally measure the bulk motion of kinetic space-time are therefore proposed here, although their practicality remains to be demonstrated.

#### A. Determination of Kinetic Bulk Motion

The kinetic theory agrees with the special theory of relativity, in that no reference frame is "preferred", that physics is the same in all inertial frames. However, it suggests that the reference frame of the bulk motion of gyrons could be detected with respect to any other intertial frame.

One method of detection uses precise clocks and the effect of time dilation. An observer in a box can determine his motion with respect to the inertial kinetic space frame by observing time dilation relative to such motion. First light pulses must be sent out in opposite directions to center the observer in the box with respect to two opposite walls, and to set synchronized clocks at these walls, using Einstein's synchronization method. The distance from the observer to either wall is $L$. The observer then sends two other clocks (A and B) away at a speed $v$ (in the box frame) towards the walls, such that clock A moves in the same direction as the bulk kinetic flow, and clock B opposite to it. A pure relativistic approach (no motion relative to kinetic inertial frame) would predict that each clock A and B will read an identical dilated time upon arrival at the edges of the box, which will differ from the time difference in the box frame ($\Delta t = L/v$):

$$\Delta t_A = \Delta t_B = \Delta t\left(1 - v^2/c^2\right)^{1/2} \qquad (23)$$

In the kinetic theory of space-time the dilation factors will be different for each clock:

$$\Delta t_A = \Delta t\left(\frac{1-(U+v)^2/c^2}{1-U^2/c^2}\right)^{1/2} \quad \text{and} \quad \Delta t_B = \Delta t\left(\frac{1-(U-v)^2/c^2}{1-U^2/c^2}\right)^{1/2} \qquad (24)$$

which can be calculated by first comparing the synchronized clock time $t$ with the time of a clock in the kinetic inertial frame and then comparing the moving clocks A and B with that frame as well. $U = \bar{v}$ is the speed of the box and observer relative to the kinetic inertial frame. In the limit of small motion of the box with respect to the moving clocks ($U/c \ll v/c$) the usual form (23) is recovered. The first order difference between the two clock measurements will be: $\Delta t_B - \Delta t_A \cong 2Uv/c^2$.

These time dilations are compatible with previous experiments using atomic clocks in moving airplanes [Hafele & Keating, 1972], which measured different time dilations for the planes moving westward and eastward around the Earth due to the rotation of the Earth. The above analysis assumes an infinite impulse imparted to the two clocks A and B from the center, whereas a real experiment might need to take into account an acceleration and/or a deceleration. This would make the time dilations (22) and (23) more complicated, but would not change the fundamental difference. Such an experiment could also be performed using decaying particles with known half lives, with flux detectors at the "edges of the box" rather than synchronized clocks to determine the relative motion.

### B. Bulk Velocity and Magnetic Vector Potential

Another prediction of the kinetic theory is that the magnetic vector potential is proportional to the bulk speed of the spatial constituents. Such a proportionality could be measured, i.e. with the above atomic clock experiment, and a vector potential imposed through the "box", or with a modified Sagnac interferometer. A Sagnac interferometer sends two light beams around a closed loop in opposite directions and measures differences in travel times with a phase shift and resulting interference between the two beams. The kinetic theory explains the Sagnac effect, which enables precise optical gyroscopes as used in aerospace technology, in terms of the rotation of the gyroscope (or experimental apparatus) with respect to the inertial frame of the hypothetical kinetic bulk flow. The famous Michelson-Gale experiment [Michelson & Gale, 1925], which showed a displacement dependent on the rotational velocity vector of the Earth, is also consistent with this explanation.

If a Sagnac interferometer were constructed with a magnetic vector potential aligned with the light beam, an interference fringe shift could be observed as the light takes longer to traverse the loop against the bulk flow. Such an experiment could be implemented with a square or rectangular interferometer, and a vector potential aligned along one leg of the apparatus by using a solenoidal current. A solenoid would limit the magnetic field and thus Faraday rotation of the beam, leaving only the vector potential to act on the light beams. With a long enough interferometer and a strong enough potential, and/or a second oppositely oriented solenoid on the opposite leg of the interferometer, a fringe shift should be observable as current is turned on through the solenoid.

## IV. CONCLUSIONS

While a kinetic description of constituents having only translational freedom is extremely powerful and suffices to describe many statistical ensembles, consideration of a spin component is sometimes beneficial as new properties will emerge. In particular, a wave of fluctuating bulk spin is postulated, relying on rotational freedom of fundamental constituents. Such a wave should gain importance relative to sound-like disturbances as kinetic media approach the limit of incompressibility.

The equations of motion, taken in conjunction with a kinetic theory of space-time, can lead not only to a non-relativistic Schrodinger equation (compressive) under certain assumptions but also to incompressive light waves, Maxwell's equations, and the Lorentz force. The equations of state that generate the wave equation allow a perturbation due to the presence of matter that leads to an interpretation of gravitational field and the dynamic metric tensor of General Relativity. Experimentally verifiable predictions are made based on the kinetic theory.

The theory as stated leaves many intriguing possibilities, which cannot all be printed here. What force could arise from a non-zero collision tensor $G_{ij}$ or a non-zero spin pressure tensor? What is the exact form of the stationary wave that gives rise to the electron or other fundamental particles in free space? If the electric potential $\phi$ is defined from an isotropic pressure tensor, what are the perturbative effects on electromagnetic fields in an anisotropic medium such as near a strong gravitational source? And could a measurable divergence of $\bar{B}$ exist on small timescales, where the bulk spin coupling to vorticity is not instantaneous? More work is needed to answer these questions and establish the mathematical tools needed to discuss the equations of state in a ten (or more) dimensional phase space.

# REFERENCES

[1] C. F. Curtiss, "Kinetic Theory of Non-spherical molecules" J. Chem. Phys., **24**, 225; 1956

[2] F. Meno, "A Plank-Length Atomistic Kinetic Model of Physical Reality", Phys. Essays, **4**, 94, 1991

[3] A. Barut & A. Braken, "Zitterbewegung and the internal geometry of the electron", Phys. Rev. D, **23**, 2454, 1981

[4] E. Madelung, Zeitschr.fr Phys. **40**, 327, 1926

[5] F. Winterburg, "Derivation of quantum mechanics from the Boltzmann equation for the Planck aether", International Journal of Theoretical Physics, **34**, 2145, 1995

[6] E. Recami & G. Salesi, "Kinematics and Hydrodynamics of Spinning Particles", Phys. Rev. A, **57**, 98-105, 1998

[7] G. Kaniadakis, "Statistical Origin of Quantum Mechanics", Physica A **307**, 172; 2002

[8] Marmanis, H., "Analogy between the Navier-Stokes and Maxwell's equations: Application to Turbulence", Phys. Fluids 10 (6), 1428-1437, 1998

[9] A. Einstein, "On the Electrodynamics of Moving Bodies", Annalen der Physik, 1905

[10] P. A. M Dirac, "General Theory of Relativity", Wiley, NY, 1975

[11] J.C. Hafele and R.E. Keating, "Around-the-world atomic clocks: predicted [& observed] relativistic time gains", Science, Vol.177, p.166, 1972

[12] Michelson, M & Gale, H, "The Effect of Earth's Rotation on the Velocity of Light", Astrophysical Journal, vol. 61, p.140, 1925

# ERRATA

In section *3A* the time dilation caused by motion through a kinetic media is calculated from two reference frames. There is an error in this calculation as the proper formula for relativistic addition or subtraction of velocities was not used. Proper attention was not paid to defining a measured velocity in a clock frame in terms of measurements of that clock. In equations (24), the quantities (U+V) and (U-V) which appear should be replaced with:

$$\frac{U+v}{(1+Uv/c^2)}$$

This factor cancels out any difference in the measured time dilations of the two clocks, and so no motion of the space-time medium can be observed, a well known result in the Lorentz-Fitzgerald relativity theory, resulting from Poincare invariance. The formulae for bulk spin waves in a kinetic fluid as presented and the other experimental prediction are not affected by this error.